# Coupling emission from single localized defects in 2D semiconductor to surface plasmon polaritons

*Tao Cai,[a] Subhojit Dutta,[a] Shahriar Aghaeimeibodi,[a] Zhili Yang,[a] Sanghee Nah,[b] John T. Fourkas,[b,c] and Edo Waks[a,d],\**

[a] Department of Electrical and Computer Engineering and Institute for Research in Electronics and Applied Physics, University of Maryland, College Park, Maryland 20742, USA.

[b] Department of Chemistry and Biochemistry, University of Maryland, College Park, Maryland 20742, USA

[c] Institute for Physical Science and Technology, University of Maryland, College Park, Maryland 20742, USA

[d] Joint Quantum Institute, University of Maryland and the National Institute of Standards and Technology, College Park, Maryland 20742, USA.

\* Corresponding author E-mail: edowaks@umd.edu

**ABSTRACT**: Coupling of an atom-like emitter to surface plasmons provides a path toward significant optical nonlinearity, which is essential in quantum information processing and quantum networks. A large coupling strength requires nanometer-scale positioning accuracy of the emitter near the surface of the plasmonic structure, which is challenging. We demonstrate the coupling of single localized defects in a tungsten diselenide (WSe$_2$) monolayer self-aligned to the surface plasmon mode of a silver nanowire. The silver nanowire induces a strain gradient on the monolayer at the overlapping area, leading to the formation of localized defect emission sites that are intrinsically close to the surface plasmon. We measure a coupling efficiency with a lower bound of 39% from the emitter into the plasmonic mode of the silver nanowire. This technique offers a way to achieve efficient coupling between plasmonic structures and localized defects of 2D semiconductors.

**KEYWORDS**: Surface plasmon polaritons, Silver nanowire, 2D semiconductors, Transition metal dichalcogenide, Tungsten Diselenide, Defect emission

Surface plasmons at the interface between a conductor and a dielectric concentrate light to subwavelength dimensions.[1,2] An atom-like emitter placed in a high-field region of a surface plasmon mode exhibits large Purcell enhancement,[3-8] and strong optical nonlinearity.[9-11] These phenomena are essential in applications such as quantum information processing[12] and quantum networking.[13]

The field of surface plasmon polaritons decays within a few tens of nanometers from the surface,[1] so strong interactions require nanometer-scale alignment of the emitter. To date, most of the reported coupled systems have relied on the random deposition of emitters around plasmonic structures.[3-6] This approach has the potential to create many devices, but at a low yield of successful devices. Deterministic fabrication techniques have also been explored, such as using an atomic force microscope (AFM) cantilever[7] or the glass fiber tip of a shear-force microscope[8] to control the position of an emitter relative to the plasmonic structure. These methods have a high yield of successful devices, but require complex setups and create devices one at a time, so are difficult to scale up. A random deposition approach with high yield of successful devices is highly desirable, and could significantly improve scalability.

Recently, bound excitons localized by defects in two-dimensional (2D) semiconductors[14] have emerged as a new class of atom-like emitters that can exhibit high-quantum-yield, single-photon emission.[15-21] These localized excitons also hold a number of advantages for coupling to surface plasmons. First, unlike quantum dots and defects in solids that are embedded in a dielectric matrix, these emitters reside in a 2D sheet. This allows the excitons to come much closer to the surface of the conductor. Second, strain engineering enables the formation of these defects at desired locations.[22-25] Third, these emitters have a relatively narrow linewidths (~100 μeV),[15-18] which offers the possibility of achieving strong coupling to plasmonic structures.[10,11] Finally, these emitters can serve as single qubits for carrying quantum information, using both electron spins and valley pseudospins inherited from the 2D semiconductor band structure.[26,27] Nevertheless, achieving efficient coupling between single defects in 2D semiconductors and plasmonic nanostructures remains challenging.

Here we demonstrate that single defects in 2D semiconductors can efficiently couple to surface plasmon polaritons. We deposited atomically thin $WSe_2$ sheets on colloidally synthesized silver nanowires. Localized, atom-like defects naturally form along the nanowire surface due to an induced strain gradient. Due to the proximity of the induced defects to the surface, they efficiently couple to propagating surface plasmon polaritons. We show that at least 39% of the emission from a single defect couples to the nanowire and scatters from the ends. Such a coupled system could be used for applications such as ultrafast single-photon sources,[4,6] which paves a way toward super-compact plasmonic circuits.

Figure 1(a) shows a schematic of the device. We first deposited chemically synthesized, bi-crystalline silver nanowires,[28,29] with average diameter of 100 nm and lengths varying from 3 µm to 10 µm, on a glass slide. Figure 1(b) shows a scanning electron micrograph (SEM) of the silver nanowires used in the experiment. We employed a dilute enough solution that the deposited nanowires were well separated, enabling us to isolate individual nanowires on an optical microscope. After depositing nanowires, we transferred a monolayer of $WSe_2$ sheets on top of the nanowires. Figure 1(c) shows a microscopic image of the $WSe_2$ monolayers used in the experiment. We synthesized these monolayers on a sapphire substrate using a chemical vapor deposition method,[30] then transferred them to the glass slide using a polydimethylsiloxane (PDMS) substrate as an intermediate transfer medium.[31] We grew a high density of monolayer flakes on the substrate to ensure that the $WSe_2$ covered a large fraction of the nanowires after transfer.

To characterize the sample, we cooled it to 3.2K in an attoDRY cryostat (Attocube Inc.). We performed photoluminescence measurements using a confocal microscope. To excite the $WSe_2$ monolayer, we focused the laser on the sample surface using an objective lens with a numerical aperture of 0.8. By adjusting the collimation of the input laser we attained either a small, diffraction-limited spot that was used to excite a specific point on a nanowire, or a larger spot that was used to excite the entire length of the nanowire. We excited the sample using either a continuous-wave laser emitting at 532 nm, or a mode-locked Ti:sapphire laser emitting at 710 nm, with a 2 ps pulse duration and a 76 MHz repetition rate (Mira 900, Coherent, Inc.). We used the same objective lens to collect the photoluminescence from the sample. Pump laser light was rejected using a long-pass optical filter. We used a half-wave plate and a polarizing beam splitter to direct the collected signal either to a monochrome scientific camera (Rolera-XR, Qimaging, Inc.) for imaging, or to a single-mode fiber that acted as a spatial filter. The single-mode fiber was used to deliver the signal to a grating spectrometer (SP2750, Princeton Instruments). A flip mirror could be used to direct the grating-resolved signal to a CCD camera (PyLoN100BR, Princeton Instrument) to measure emission spectra, or to a single-photon-counting avalanche diode (Micro Photon Devices) to perform time-resolved measurements.

Figure 2(a) shows a photoluminescence intensity map of a sample excited with the continuous-wave laser. The red box indicates the position of the silver nanowire. We used an un-collimated laser to generate a large focal spot that excited a large portion of the silver nanowire. In addition to diffuse photoluminescence from the entire excitation region, the image reveals bright localized emission spots along the length of the nanowire. Such localized emission has been reported to be signature of single localized emitters.[15-18] We find that these defects formed preferentially where the $WSe_2$ monolayer covered the nanowire. We observed this behavior in numerous nanowires (see supporting information, supplementary figure 1). We attribute the formation of defects to strain induced by the nanowire. Similar strain-driven defect formation has been reported for patterned substrates that contained holes[22] or micropillars.[25]

To verify that the bright localized spots are single, atom-like emitters, we measured their emission spectrum. We used a large focal spot of the continuous-wave laser to excite multiple emitters along the wire and collected their photoluminescence. Figure 2(b) shows the resulting photoluminescence spectrum, which exhibits several sharp emission lines. The spectrum from the region containing the localized defects is distinct from the spectrum collected at a bare $WSe_2$ monolayer area, which displays two broad peaks corresponding to the exciton and ensemble of defects and impurities of the monolayer (Figure 2(c)). The sharp spectral emission at the bright localized regions supports the assertion that these spots are single, strain-induced emitters.

Figure 3(a) shows the spectrum of a representative single emitter along a nanowire. The emitter was excited using the continuous-wave laser with a highly focused spot, and the spectrum was obtained over a narrow spectral range. The spectrum is a doublet, each peak of which corresponds to one of two orthogonal linear polarizations of the emission. A fit of the spectrum to two Lorentzian functions gives linewidths of 0.08 nm (183 μeV) for the peak at 736.74 nm and 0.16 nm (365 μeV) for the peak at 737.07 nm, with a splitting of 0.33 nm (745 μeV). This spectrum is similar to that of those natural, defect-bound, localized emitters observed previously in $WSe_2$ monolayers.[15-18]

Figure 3(b) shows the integrated intensity of the photoluminescence of the same emitter in Figure 3(a) as a function of excitation power of the continuous-wave laser. The saturation behavior observed is consistent with single-defect emission. The solid line is a fit to a saturation function of the form $I = I_{sat}P/(P_{sat} + P)$, where $I$ and $I_{sat}$ are the integrated intensity and the saturation intensity, respectively, and $P$ and $P_{sat}$ are excitation power and saturation power, respectively. From the fit we determined a power of 3.6 W/cm² (before the objective lens) and an integrated intensity of $3\times10^4$ counts/s on the spectrometer at saturation.

For further confirmation that the emission arises from a defect, we performed time-resolved photoluminescence measurement on the emitter in Figure 3(a). We excited the emitter with pulses of the Ti:sapphire laser, and measured the photoluminescence using a single-photon-counting avalanche diode. A time-correlated, single-photon counting system (PicoHarp 300, PicoQuant Inc.) was used to determine the signal decay of the emitter as a function of time following the laser pulse. Figure 3(c) shows the time-resolved photoluminescence of the emitter. We normalized the curve with respect to its count rate at time $t = 0$. We fit the decay to a single exponential (solid line) with a lifetime of 2.4 ns. This lifetime is significantly longer than that of the excitonic emission of an WSe$_2$ monolayer, which is typically on the order of a few picoseconds at a temperatures near 4 K.[32,33] However, the measured lifetime is consistent with that of a confined quantum emitter.[34-37] The signal in Figure 3(c) does not decay to zero at $t \gg 0$, which is due to non-zero count on the avalanche diode as a result of environment light and dark noise at all times.

One fortuitous advantage of the strain-driven formation process is that the emitters are naturally generated near the high-field region of the wire. We therefore expect the emitters to couple efficiently to guided surface plasmon modes. Figure 4(a) shows a photoluminescence intensity map of an emitter near the midpoint of a silver nanowire. We focused the continuous-wave laser on the emitter at the location denoted "C". In addition to the photoluminescence from the emitter itself, we also observed emission from both ends of the silver nanowire (denoted "A" and "B"). We attribute the light at the ends of the nanowire to emission from the emitter that couples to guided surface plasmon polaritons and travels to the ends. To verify that the emission at the wire ends originates from the defect, we measured the emission spectrum at all three points (Figure 4(b)). All three emission spectra are identical, which demonstrates that the emission originates from the same source. Furthermore, when we moved the excitation spot away from the emitter, all three peaks disappeared.

We can estimate the coupling efficiency of the defect emission into the plasmonic mode of the silver nanowire by comparing the emission intensity at both ends of the nanowire to the overall emission of the emitter. The coupling efficiency is given by

$$\beta = (I_A + I_B + L)/(I_A + I_B + I_C + L) \qquad (1)$$

in which $I_i$ ($i = A, B$ or $C$) is the integrated intensity at location $i$ and $L$ is number of photons lost while propagating along the silver nanowire. We do not have an accurate means of measuring $L$, and therefore cannot calculate the exact coupling efficiency. However, $\beta_{min} = \frac{I_A + I_B}{I_A + I_B + I_C}$ is a lower bound to the coupling efficiency, which in this case is found to be 39%. $\beta_{min}$ is quite high, which indicates a substantial coupling efficiency between the surface plasmon mode and the nanowire. We note that our calculation assumes that the collection efficiency at all three points is approximately equal. This approximation is reasonable, because each emission spot originates from a sub-wavelength source.

In summary, we have demonstrated efficient coupling between a localized emitter in atomically thin $WSe_2$ and the propagating surface plasmon mode of a silver nanowire. The nanowire induces a strain gradient that results in emitters being formed naturally along the nanowire surface. We measured a lower-bound coupling efficiency of 39% from the emitter into the plasmonic mode of the silver nanowire. Although the technique we presented here does not enable a deterministic positioning of emitters near the plasmonic structure, it does induce emitters self-aligned to the surface plasmon polaritons with efficient coupling. This could lead to a high yield of coupled devices as long as the atomically thin $WSe_2$ sheet appropriately covers a large number of plasmonic structures. To gain a control over the coupling strength between the emitter and the surface plasmon polariton, one potential path would be to insert a buffer layer, such as a boron nitride of controlled thickness,[38] between the monolayer $WSe_2$ and the silver nanowire. With a large enough coupling strength between the emitter and surface plasmon polariton, it is possible to obtain a coupled system with nonlinearity at the single-photon level, which is essential for applications such as photonic transistors.[10] The technique presented here is versatile, and could be applied to construct coupled systems consisting of diverse plasmonic structures[39] and single defects in a range of 2D semiconductors.[14-21]

## ASSOCIATED CONTENT

**Supporting information**

The following files are available free of charge: Photoluminescence intensity maps of a $WSe_2$ monolayer over a silver nanowire (PDF)

## AUTHOR INFORMATION

**Corresponding Author**

*E-mail: edowaks@umd.edu.

**Funding Sources**

The authors acknowledge support from the National Science Foundation (award number ECCS1508897), the Office of Naval Research ONR (award number N000141410612), the Air Force Office of Scientific Research (AFOSR) (award number 271470871D), and the Physics Frontier Center at the Joint Quantum Institute.

**Notes**

The authors declare no competing financial interest.

FIGURES

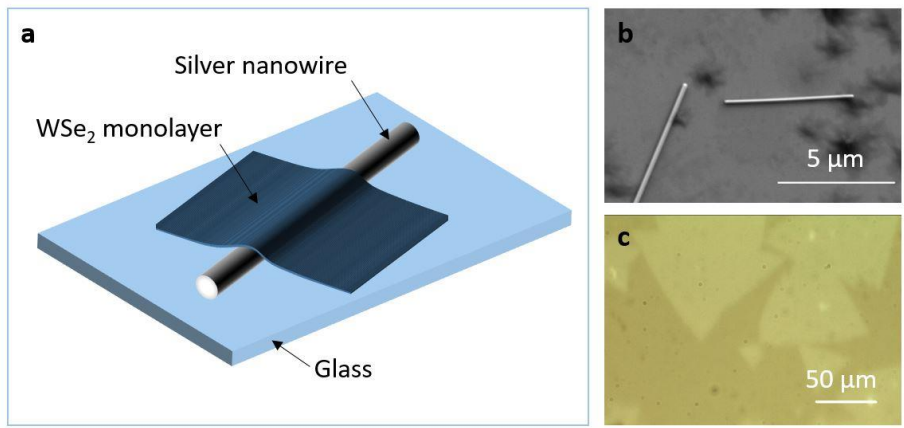

Figure. 1. (a) 3D schematic layout of a silver nanowire/WSe$_2$ monolayer device. (b) Scanning electron micrograph showing the silver nanowires used in the experiment. (c) Optical micrograph of WSe$_2$ monolayer flakes grown by chemical vapor deposition. Lighter areas correspond to the monolayer.

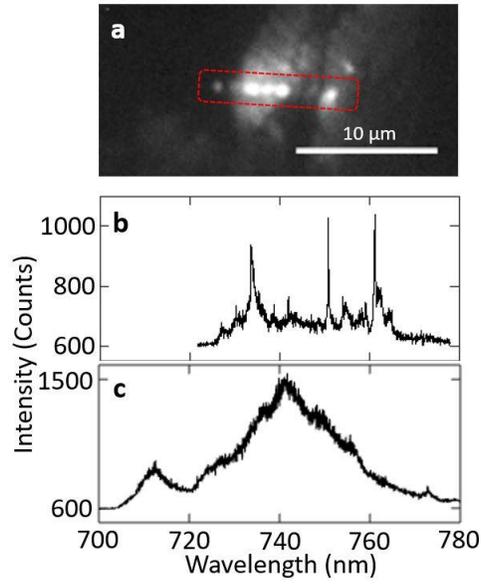

Figure. 2. (a) Photoluminescence intensity map of a WSe$_2$ monolayer over a silver nanowire. The red box indicates the position of the silver nanowire. (b) Photoluminescence spectrum of the bright localized spots along the silver nanowire. (c) Photoluminescence spectrum of a bare WSe$_2$ monolayer.

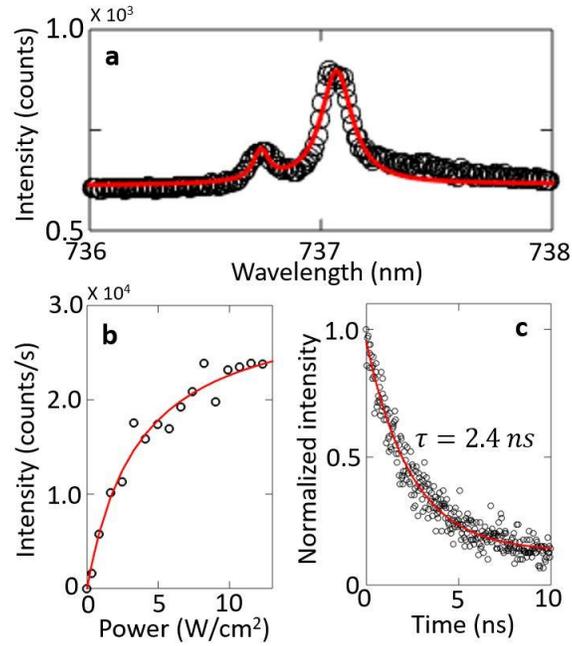

Figure. 3. (a) Photoluminescence spectrum of a localized emitter (black circles) fitted with two Lorentzian functions (red curve) (b) The integrated photoluminescence intensity of a single emitter as a function of excitation power (black circles) shows a saturation behavior (red curve). (c) The time-resolved photoluminescence (black circles) of a single emitter can be fit to an exponential decay function (red curve).

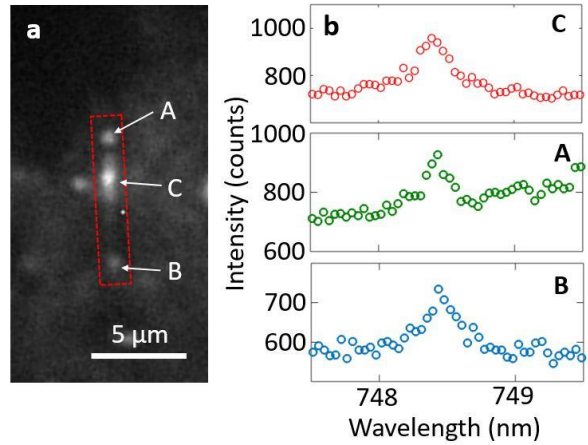

Figure. 4. (a) A photoluminescence intensity map shows emission at the emitter (denoted "C") and at both ends of the silver nanowire (denoted "A" and "B"). The red box indicates the position of the silver nanowire. (b) Photoluminescence spectra collected at "C" (red circles), "A" (green circles) and "B" (blue circles).